\documentclass[preprintnumbers,article,amsmath,amssymb,floatfix,10pt,prd,twocolumn,superscriptaddress,nofootinbib]{revtex4-2}
\usepackage{bm}
\usepackage{amsfonts}
\usepackage{latexsym}
\usepackage[utf8]{inputenc}
\usepackage{graphicx}
\usepackage{amsmath}
\usepackage{palatino}
\usepackage{mathpazo}
\usepackage{textcomp}
\usepackage{float}
\usepackage{booktabs}
\usepackage{dcolumn}
\usepackage{ragged2e}
\usepackage{hyperref}
\hypersetup{colorlinks,citecolor=blue}
\hypersetup{colorlinks=true,linkcolor=blue,filecolor=magenta,urlcolor=blue}
\usepackage{amsmath}
\usepackage{subfigure}
\usepackage[]{natbib}
\usepackage{xcolor}
\usepackage{orcidlink}
\usepackage{epsfig}
\usepackage{caption}
\usepackage[toc]{appendix}
\usepackage{commath}
\usepackage{cancel}
\usepackage{csquotes}
\usepackage{placeins}

\usepackage{multirow}

\def\jnl@style{\it}
\def\aaref@jnl#1{{\jnl@style#1}}

\def\aaref@jnl#1{{\jnl@style#1}}

\def\aj{\aaref@jnl{AJ}}                   
\def\apj{\aaref@jnl{ApJ}}                 
\def\apjl{\aaref@jnl{ApJ}}                
\def\apjs{\aaref@jnl{ApJS}}               
\def\apss{\aaref@jnl{Ap\&SS}}             
\def\aap{\aaref@jnl{A\&A}}                
\def\aapr{\aaref@jnl{A\&A~Rev.}}          
\def\aaps{\aaref@jnl{A\&AS}}              
\def\mnras{\aaref@jnl{Mon.~Not.~Roy.~Astron.~Soc.}}             
\def\prd{\aaref@jnl{Phys.~Rev.~D}}        
\def\prc{\aaref@jnl{Phys.~Rev.~C}}  
\def\prl{\aaref@jnl{Phys.~Rev.~Lett.}}    
\def\qjras{\aaref@jnl{QJRAS}}             
\def\skytel{\aaref@jnl{S\&T}}             
\def\ssr{\aaref@jnl{Space~Sci.~Rev.}}     
\def\zap{\aaref@jnl{ZAp}}                 
\def\nat{\aaref@jnl{Nature}}              
\def\aplett{\aaref@jnl{Astrophys.~Lett.}} 
\def\apspr{\aaref@jnl{Astrophys.~Space~Phys.~Res.}} 
\def\physrep{\aaref@jnl{Phys.~Rep.}}      
\def\physscr{\aaref@jnl{Phys.~Scr}}       
\def\commat{\aaref@jnl{Comm.~Math.~Phys.}}              
\def\science{\aaref@jnl{Science}}               
\def\cqg{\aaref@jnl{Classical Quant.~Grav.}}            
\def\jpcs{\aaref@jnl{JPCS}}                                     
\def\ijmpd{\aaref@jnl{Int.~J.~Mod.~Phys.~D}}                    
\def\grg{\aaref@jnl{Gen.~Relat.~Gravit.}}               
\def\rpp{\aaref@jnl{Rep.~Prog.~Phys.}}          
\def\npa{\aaref@jnl{Nucl.~Phys.~A}}        
\def\lrr{\aaref@jnl{Living Rev.~Rel.}}                   
\def\jcap{\aaref@jnl{J.~Cosmology Astropart.~Phys.}}    
\def\rmp{\aaref@jnl{Rev.~Mod.~Phys.}}   
\def\epjc{\aaref@jnl{Eur.~Phys.~J.~C}}
\def\plb{\aaref@jnl{~Phy.~Lett.~B}}
\def\mpla{\aaref@jnl{Mod.~Phy.~Lett.~A}}
\def\arxiv{\aaref@jnl{arxiv.org}}

\allowdisplaybreaks[1]

\begin{document}

\color{blue}

\title{Constraining $f(Q, L_m)$ gravity with bulk viscosity}

\author{Y. Myrzakulov\orcidlink{0000-0003-0160-0422}}\email[Email: ]{ymyrzakulov@gmail.com} 
\affiliation{Department of General \& Theoretical Physics, L.N. Gumilyov Eurasian National University, Astana, 010008, Kazakhstan.}

\author{O. Donmez\orcidlink{0000-0001-9017-2452}}
\email[Email: ]{orhan.donmez@aum.edu.kw}
\affiliation{College of Engineering and Technology, American University of the Middle East, Egaila 54200, Kuwait.} 

\author{M. Koussour\orcidlink{0000-0002-4188-0572}}
\email[Email: ]{pr.mouhssine@gmail.com}
\affiliation{Department of Physics, University of Hassan II Casablanca, Morocco.} 

\author{S. Muminov\orcidlink{0000-0003-2471-4836}}
\email[Email: ]{sokhibjan.muminov@gmail.com}
\affiliation{Mamun University, Bolkhovuz Street 2, Khiva 220900, Uzbekistan.}

\author{I. Y. Davletov}
\email[Email: ]{ikram.d@urdu.uz}
\affiliation{Technical Faculty, Urganch State University, Urgench, Uzbekistan.}

\author{J. Rayimbaev\orcidlink{0000-0001-9293-1838}}
\email[Email: ]{javlon@astrin.uz}
\affiliation{Institute of Fundamental and Applied Research, National Research University TIIAME, Kori Niyoziy 39, Tashkent 100000, Uzbekistan.}
\affiliation{University of Tashkent for Applied Sciences, Str. Gavhar 1, Tashkent 100149, Uzbekistan.}
\affiliation{Shahrisabz State Pedagogical Institute, Shahrisabz Str. 10, Shahrisabz 181301, Uzbekistan.}
\affiliation{Tashkent State Technical University, Tashkent 100095, Uzbekistan.}

\begin{abstract}
We investigate the influence of bulk viscosity on late-time cosmic acceleration within an extended $f(Q, L_m)$ gravity framework, where the non-metricity $Q$ is non-minimally coupled with the matter Lagrangian $L_m$. Analyzing the function $f(Q, L_m) = \alpha Q + \beta L_m$, we derive exact solutions under non-relativistic matter domination. Using observational datasets ($H(z)$, Pantheon supernovae, and their combination), we constrain the model parameters $H_0$, $\alpha$, $\beta$, and $\zeta$. The deceleration parameter transitions from positive to negative values around redshifts $z_t \approx 0.80$ to $0.99 $, indicating current accelerated expansion. Moreover, the effective equation of state parameter, $\omega_{eff}$, resembles quintessence dark energy ($-1 < \omega_{eff} < -\frac{1}{3}$), with corresponding values from respective datasets. Finally, we use the $Om(z)$ diagnostic, which confirms that our model demonstrates quintessence-like behavior. Our findings underscore the significant role of bulk viscosity in understanding accelerated expansion in the universe within alternative gravity theories.

\end{abstract}

\maketitle


\section{Introduction}\label{sec1}

Recent observations of type Ia supernovae (SNe Ia) have shown that the universe is expanding at an accelerated rate \cite{Riess/1998,Riess/2004,Perlmutter/1999}. To explain this acceleration, physicists have proposed the existence of dark energy (DE), a fluid with sufficient negative pressure, which constitutes about 70\% of the universe's total energy and matter content. The simplest explanation for DE is the introduction of a cosmological constant (CC), which accounts for the observed accelerated expansion and forms the basis of the $\Lambda$CDM model, a model that has been highly successful \cite{Zlatev/1999}. Despite the observational support for the CC model, it faces several significant issues. One major problem is the substantial discrepancy between the theoretically predicted and observed values of the CC. Another issue is the cosmic coincidence problem, which refers to the fact that we appear to live in a universe where the matter density and DE density are of the same order. These challenges motivate researchers to explore alternative models for the universe's accelerated expansion \cite{Weinberg/1989,Padmanabhan/2003,Steinhardt/1999}.

Modified gravity theories (MGTs) has recently become a significant branch of modern cosmology, aiming to offer a unified explanation for both the early universe and the accelerated expansion observed in later stages. These theories extend the geometry of Einstein's general relativity (GR) by modifying the Einstein-Hilbert (EH) action to account for cosmic acceleration. Numerous MGTs have been proposed to describe the acceleration of the universe during both its early and late phases. The $f(R)$ gravity theory, introduced in \cite{Buchdahl/1970}, is one of the most fundamental and widely studied modifications of GR. Numerous researchers have explored various aspects of $f(R)$ gravity, including its potential to drive cosmic inflation and acceleration \cite{Dunsby/2010,Carroll/2004}. Another extension of the EH action involves the non-minimal coupling between matter and geometry, leading to the $f(R, L_m)$ gravity. Harko and Lobo \cite{Harko/2010} introduced $f(R, L_m)$ gravity, where the gravitational Lagrangian is described by an arbitrary function of the Ricci scalar $R$ and the matter Lagrangian $L_m$. This theory has been explored for various astrophysical and cosmological implications \cite{Wang/2012,Goncalves/2023}. Myrzakulova et al. \cite{Myrzakulova/2024} explored the DE phenomenon within the framework of $f(R, L_m)$ cosmological models by incorporating observational constraints. Myrzakulov et al. \cite{Myrzakulov/2024} have presented a study on the linear redshift parametrization of the deceleration parameter within the framework of $f(R, L_m)$ gravity. Similarly, several other MGTs with distinct cosmological implications exist, including theories such as $f(G)$ theory \cite{Felice/2009,Bamba/2017,Goheer/2009}, $f(R, T)$ theory \cite{Harko/2011,Koussour_1/2022,Koussour_2/2022,Myrzakulov/2023}, $f(\mathcal{T}, B)$ theory \cite{Bahamonde/2018}, and others.

GR is formulated within the framework of Riemannian geometry. Therefore, exploring more general geometric structures that can describe the gravitational field at the solar system level presents an intriguing approach for developing extended theories of gravity. Consequently, a more coherent theory equivalent to GR, known as the teleparallel equivalent of GR (TEGR) or $f(\mathcal{T})$ theory \cite{Ferraro/2007,Myrzakulov/2011,Capozziello/2011}, has been proposed. In this framework, $\mathcal{T}$ represents the torsion, which is defined as the antisymmetric part of the connection that describes gravitational effects. Furthermore, Weyl suggested extending Riemannian geometry, which led to the development of the first combined theory of gravity and electromagnetic, in which the electromagnetic field is produced by spacetime's non-metricity. Consequently, the symmetric teleparallel representation emerges as the third generalization of GR (STEGR). This generalization extends to $f(Q)$ gravity \cite{Jimenez/2018}, where the fundamental geometric variable is the non-metricity $Q$. This non-metricity geometrically defines the variation in the length of a vector during parallel transport, characterizing the properties of gravitational interaction. Despite being recently proposed, $f(Q)$ gravity theory already showcases several intriguing and useful applications in the literature. Refs. \cite{Jimenez/2020,Khyllep/2021} include the first cosmological solutions in $f(Q)$ gravity, whereas \cite{MK1, MK2, MK3,MK4,MK5,MK6,MK7,MK8} address cosmic acceleration and DE. Moreover, Xu et al. \cite{Xu/2019,Xu/2020} introduced an extension of $f(Q)$ theory, termed $f(Q, T)$, wherein the non-metricity scalar $Q$ is non-minimally coupled to the trace $T$ of the energy-momentum tensor. Early investigations by Xu et al. \cite{Xu/2019} initially examined the cosmological implications of $f(Q, T)$ theory. Later research \cite{K6,K7} focused on exploring late-time accelerated expansion within observational constraints. In addition, other areas such as baryogenesis mechanism \cite{Bhattacharjee}, inflation \cite{Shiravand}, and perturbations \cite{Najera} have received significant attention. Nonetheless, there has been relatively little research dedicated to studying the astrophysical consequences of $f(Q, T)$ theory \cite{Tayde1,Sneha2,Bourakadi}. Building upon $f(R, L_m)$ theory, Hazarika et al. \cite{Hazarika/2024} extended STEGR by integrating the matter Lagrangian into the Lagrangian density of $f(Q)$ theory, thereby formulating $f(Q, L_m)$ theory. This framework allows for minimal and non-minimal couplings between geometry and matter. The authors derived the general system of field equations by varying the action with respect to the metric. They also investigated the conservation of the matter energy-momentum tensor and demonstrated that it is not conserved in this theory. In addition, they explored the cosmic evolution under a flat Friedmann-Lema\^{i}tre-Robertson-Walker (FLRW) metric, deriving the generalized Friedmann equations. Furthermore, the authors examined two specific gravitational models corresponding to distinct forms of the function $f(Q, L_m)$, namely $f(Q, L_m) = -\alpha Q + 2 L_m + \beta$ and $f(Q, L_m) = -\alpha Q + (2 L_m)^2 + \beta$ \cite{Hazarika/2024}. 

In this paper, we study the dominance of bulk viscous matter in the context of the newly modified gravity theory $f(Q, L_m)$. Viscosity has been identified as a crucial factor in understanding the universe's evolution. Various dissipative processes \cite{Misner/1968,Zimdahl/2001} in the early universe lead to deviations from the perfect fluid assumptions \cite{Pavon/1991,Lima/1996}, allowing for the presence of viscosity \cite{Brevik1/2017,Brevik/2018}. Consequently, exploring cosmological solutions that include viscosity is crucial for understanding the different phases of the universe's evolution \cite{Brevik2/2017,Elizalde/2018}. In the early universe, bulk viscosity can arise from various physical processes, including the decoupling of matter from radiation during the recombination era, particle collisions involving gravitons, and the formation of galaxies \cite{Barrow/1977}. These processes can influence the expansion rate and play a role in the evolution of the universe at different stages, from the radiation-dominated phase to later epochs. As Weinberg \cite{bulk1} discussed, bulk viscosity contributes to the entropy production in the universe. In the early universe, bulk viscosity could have served as a dissipative mechanism that smoothed out initial anisotropies. This process could explain the high entropy per baryon ratio that we observe today, which suggests that bulk viscosity played a crucial role in the isotropization of the universe during its radiation-dominated phase \cite{Misner/1968}. Moreover, Murphy's work \cite{bulk2} has shown that bulk viscosity may prevent the appearance of the Big Bang singularity, which indicates that it could modify the universe's behavior at very early times. The influence of viscosity on cosmological evolution and its potential role in avoiding the initial singularity has been explored by several researchers \cite{bulk3,bulk4,bulk5,bulk55}. Other studies \cite{bulk6,bulk7} have connected bulk viscosity to grand unified theory (GUT) scenarios, suggesting that it might be involved in driving inflationary behavior. During inflation, bulk viscosity could have provided a phenomenological description of quantum particle creation in strong gravitational fields, further influencing the universe's expansion dynamics in its earliest moments. Observations also suggest that significant bulk viscous stress plays an important role in the late-time evolution of the universe \cite{Pavon/1993,Nojiri/2005}. Eckart \cite{Eckart/1940} was the first to formulate a relativistic theory of viscosity. However, Eckart's theory has notable shortcomings, particularly regarding causality and stability \cite{Hiscock/1986}. To address these issues, Israel and Stewart \cite{Israel/1979} developed a fully relativistic formulation known as extended irreversible thermodynamics, which offers a satisfactory alternative to Eckart's theory. By taking into account the effect of the bulk viscosity coefficient $\zeta$ in the usual cosmic pressure, we analyze the impact on the universe's evolution phase. Using a scaling law for the viscosity coefficient $\zeta$, we reduce the Einstein case to a form proportional to the Hubble parameter \cite{Brevik/2012}. This scaling law proves to be highly useful. To review several interesting cosmological models that include the viscosity in cosmic fluid, one can go to the Refs. \cite{Brevik/2013,Brevik/2005,Mohan/2017,Koussour_fQ1,Koussour_fQ2,Koussour_fQ3,Odintsov/2020,Fabris/2006}.

The structure of this study unfolds as follows. Sec. \ref{sec2} introduces the action and foundational formulation governing the dynamics within $f(Q, L_m)$ gravity. In Sec. \ref{sec3}, we derive the field equations under the FLRW metric. In addition, we explore a cosmological $(Q, L_m)$ model and determine the Hubble parameter as a function of cosmic redshift. Sec. \ref{sec4} utilizes observational data from Hubble parameter measurements $H(z)$ and the Pantheon dataset to estimate model parameters and the bulk viscous coefficient $\zeta$. We also analyze the behavior of cosmological parameters in this section. In Sec. \ref{sec5}, we employ the $Om(z)$ diagnostic to distinguish between different DE models. Finally, Sec. \ref{sec6} concludes the discussion and summarizes our findings.

\section{$f(Q,L_{m})$ gravity theory}\label{sec2}

In Weyl–Cartan geometry, the affine connection $Y^\alpha_{\;\;\mu\nu}$ can be decomposed into three irreducible components: the symmetric Levi-Civita connection $\Gamma^\alpha_{\;\;\mu\nu}$, the contortion tensor $K^\alpha_{\;\;\mu\nu}$ representing the antisymmetric part, and the disformation tensor $L^\alpha_{\;\;\mu\nu}$, which accounts for non-metricity. Thus, it can be generally expressed as follows \cite{Xu/2019}:
\begin{equation}
Y^\alpha_{\;\;\mu\nu}=\Gamma^\alpha_{\;\;\mu\nu}+K^\alpha_{\;\;\mu\nu}+L^\alpha_{\;\;\mu\nu}.
\end{equation}

The standard description of the Levi-Civita connection for the metric $g_{\mu\nu}$, which is the first term in the equation above, is given by
\begin{equation}
    \Gamma^\alpha_{\;\;\mu\nu}=\frac12 g^{\alpha\lambda}(\partial_\mu g_{\lambda \nu}+\partial_\nu g_{\lambda \mu} - \partial_\lambda g_{\mu\nu}).
\end{equation}

The contortion tensor $K^\alpha_{\;\;\mu\nu}$ is expressed in terms of the torsion tensor $T^\alpha_{\;\;\mu\nu}$ as follows:
\begin{equation}
    K^\alpha_{\;\;\mu\nu}=\frac{1}{2}(T^\alpha_{\;\;\mu\nu}+ T_{\mu\;\;\nu}^{\;\;\alpha}+T_{\nu\;\;\mu}^{\;\;\alpha}).
\end{equation}

The contortion tensor is antisymmetric with regard to its first two indices, as can be seen from the equation above. On the other hand, the disformation tensor $L^\alpha_{\;\;\mu\nu}$ is derived from the non-metricity as follows:
\begin{equation}
    L^\alpha_{\;\;\mu\nu}=\frac{1}{2}(Q^\alpha_{\;\;\mu\nu}-Q^{\;\;\alpha}_{\mu\;\;\nu}-Q^{\;\;\alpha}_{\nu\;\;\mu}).
\end{equation}

The non-metricity tensor $ Q_{\alpha\mu\nu}$ is defined as the covariant derivative of the metric tensor with respect to the Weyl–Cartan connection $Y^\alpha_{\;\;\mu\nu}$. This relationship is given by $Q_{\alpha\mu\nu}=\nabla_\alpha g_{\mu\nu}$, and it can be derived as follows:
\begin{equation}
    Q_{\alpha\mu\nu}= \partial_\alpha g_{\mu\nu} - Y^\beta_{\;\;\alpha\mu}g_{\beta\nu}-Y^\beta_{\;\;\alpha\nu}g_{\mu\beta}.
\end{equation}

Also, we define the non-metricity conjugate, known as the superpotential $P^\alpha_{\;\;\mu\nu}$, as follows:
\begin{equation}
    P^\alpha_{\;\;\mu\nu}= -\frac{1}{2}L^\alpha_{\;\;\mu\nu}+\frac{1}{4}(Q^\alpha-\Tilde{Q}^\alpha)g_{\mu\nu}-\frac{1}{4}\delta^\alpha_{\;\;(\mu}Q_{\nu)},
\end{equation}
where $Q^\alpha=Q^{\alpha\;\;\mu}_{\;\;\mu}$ and $\Tilde{Q}^\alpha=Q_{\mu}^{\;\;\alpha\mu}$ represent the non-metricity vectors. The non-metricity scalar is obtained by contracting the superpotential tensor with the non-metricity tensor:
\begin{equation}
    Q=-Q_{\lambda\mu\nu}P^{\lambda\mu\nu}.
\end{equation}

Here, we consider an extension of symmetric teleparallel gravity, where the action is defined by \cite{Hazarika/2024}
\begin{equation}
    S=\int f(Q,L_m) \sqrt{-g} d^4x, \label{Action}
\end{equation}
where $\sqrt{-g}$ denotes the determinant of the metric, and $f(Q, L_m)$ represents an arbitrary function of the non-metricity scalar $Q$ and the matter Lagrangian $L_m$.

The gravitational field equations, which explain how spacetime geometry relates to the presence of matter and energy, are obtained via varying the gravitational action with respect to the metric tensor. Therefore, we derive the field equation for $f(Q, L_m)$ gravity,
\begin{multline}
\frac{2}{\sqrt{-g}}\nabla_\alpha(f_Q\sqrt{-g}P^\alpha_{\;\;\mu\nu}) +f_Q(P_{\mu\alpha\beta}Q_\nu^{\;\;\alpha\beta}-2Q^{\alpha\beta}_{\;\;\;\mu}P_{\alpha\beta\nu})\\
+\frac{1}{2}f g_{\mu\nu}=\frac{1}{2}f_{L_m}(g_{\mu\nu}L_m-T_{\mu\nu}),\label{field}
\end{multline}
where $f_Q=\partial f(Q,L_m)/\partial Q$ and $f_{L_m}=\partial f(Q,L_m)/\partial L_m$. For $f(Q, L_m) = f(Q) + 2\,L_m $, it simplifies to the field equation of $f(Q)$ gravity \cite{Jimenez/2018}. In addition, the energy-momentum tensor $T_{\mu\nu}$ of the matter can be expressed as 
\begin{equation}
    T_{\mu\nu}=-\frac{2}{\sqrt{-g}}\frac{\delta(\sqrt{-g}L_m)}{\delta g^{\mu\nu}}=g_{\mu\nu}L_m-2\frac{\partial L_m}{\partial g^{\mu\nu}},
\end{equation}

Again, by varying the gravitational action with respect to the connection, we derive the field equations,
\begin{equation}
    \nabla_\mu\nabla_\nu\Bigl( 4\sqrt{-g}\,f_Q\,P^{\mu\nu}_{\;\;\;\;\alpha}+H_\alpha^{\;\;\mu\nu}\Bigl)=0,
\end{equation}
where $H_\alpha^{\;\;\mu\nu}$ represents the hypermomentum density, defined as
\begin{equation}
    H_\alpha^{\;\;\mu\nu}=\sqrt{-g}f_{L_m}\frac{\delta L_m}{\delta Y^\alpha_{\;\;\mu\nu}}.
\end{equation}

By applying the covariant derivative to the field equation (\ref{field}), one can derive
\begin{multline}
D_\mu\,T^\mu_{\;\;\nu}= \frac{1}{f_{L_m}}\Bigl[ \frac{2}{\sqrt{-g}}\nabla_\alpha\nabla_\mu H_\nu^{\;\;\alpha\mu} + \nabla_\mu\,A^{\mu}_{\;\;\nu} \\ - \nabla_\mu \bigr( \frac{1}{\sqrt{-g}}\nabla_\alpha H_\nu^{\;\;\alpha\mu}\bigr) \Bigr]=B_\nu \neq 0.
\end{multline}

Therefore, in $f(Q, L_m)$ gravity theory, the matter energy-momentum tensor is not conserved. The non-conservation tensor $B_\nu $ depends on dynamical variables such as $Q$, $L_m$, and the thermodynamical quantities of the system.

\section{The cosmological model} \label{sec3}

To analyze the cosmological evolution in $f(Q,L_m)$ gravity, we study the Universe within the framework of flat FLRW geometry. This model assumes that the universe exhibits two fundamental properties \cite{ryden/2003}:
\begin{itemize}
    \item \textbf{Homogeneity}: The universe appears uniform at any given time, meaning its density and structure are the same everywhere on large scales. This uniformity suggests that there are no preferred locations within the universe.
    \item \textbf{Isotropy}: Observations from any point in the universe reveal the same physical laws and conditions in all directions. This symmetry implies that the universe looks the same when viewed from different vantage points.
\end{itemize}

In the FLRW geometry, the spacetime interval is described by the line element:
\begin{equation}
\label{FLRW}
    ds^2=-dt^2+a^2(t)(dx^2+dy^2+dz^2),
\end{equation}
where $a(t)$ is the scale factor that varies with cosmic time $t$. From metric (\ref{FLRW}), the non-metricity scalar is $Q=6 H^2$, where $H = \frac{\dot{a}}{a}$ is the Hubble parameter, which denotes the rate of expansion of the universe.

To refine the description of the cosmic fluid and minimize its idealized properties, we introduce bulk viscosity. In the context of cosmology, considering a viscous fluid is motivated by the need to account for dissipative processes that are not present in a perfect fluid. This bulk viscosity contributes negatively to the total pressure \cite{Odintsov/2020,Fabris/2006}. Due to spatial isotropy, the bulk viscous pressure is uniform in all spatial directions and is proportional to the volume expansion $\theta=3 H$. Consequently, the effective pressure of the cosmic fluid becomes \cite{Brevik/2012,Brevik/2013,Brevik/2005}
\begin{equation}
    \Bar{p}=p- \zeta \theta=p-3\zeta H,
\end{equation}
where $p$ represents the usual pressure, and $\zeta > 0$ is the bulk viscosity coefficient. In addition, the corresponding energy-momentum tensor is expressed as
\begin{equation}
    T_{\mu\nu}=(\rho+\Bar{p})u_{\mu}u_{\nu}+\Bar{p}g_{\mu\nu},
\end{equation}
where $\rho$ is the energy density, and the four-velocity of the fluid $u^\mu$ has components $u^\mu = (1, 0, 0, 0)$. The relationship between usual pressure and energy density is given by $p = (\gamma - 1)\rho$, where $\gamma$ is a constant within the range $0 \leq \gamma \leq 2$. Consequently, the effective equation of state (EoS) for the bulk viscous fluid is expressed as follows:
\begin{equation}
\label{pressure}
    \Bar{p}=(\gamma - 1)\rho-3\zeta H,
\end{equation}

The Friedmann equations, which describe the universe dominated by bulk viscous matter in $f(Q, L_m)$ gravity, are given by \cite{Hazarika/2024}
\begin{eqnarray}
\label{F1}
    && 3H^2 =\frac{1}{4f_Q}\bigr[ f - f_{L_m}(\rho + L_m) \bigl],\\
   && \dot{H} + 3H^2 + \frac{\dot{f_Q}}{f_Q}H=\frac{1}{4f_Q}\bigr[ f + f_{L_m}(\Bar{p} - L_m) \bigl]. \label{F2}
\end{eqnarray}

Specifically, when $f(Q, L_m) = f(Q) + 2\,L_m$, the Friedmann equations reduce to $f(Q)$, which can be further simplified to the symmetric teleparallel equivalent of GR.

Numerous cosmological observations and solar system tests that confirm the validity of GR indicate that any deviations from standard GR must be minimal. Consequently, the $f(Q)$ function is expected to closely follow a linear form. To explore the dynamics of the universe with viscosity, we therefore adopt the following $f(Q, L_m)$ function:
\begin{equation}
    f(Q,L_m)=\alpha Q+\beta L_m,
\end{equation}
where $\alpha$ and $\beta$ are constants. This model is inspired by the functional form $f(Q, T) = \alpha Q + \beta T$, which describes a linear coupling between matter and geometry, as explored in the literature \cite{Xu/2019, Xu/2020}. Therefore, we immediately find that $f_Q = \alpha$, and $f_{L_m} = \beta$. Thus, for this specific functional form, with $L_m = \rho$ \cite{Harko/2015}, the Friedmann equations (\ref{F1}) and (\ref{F2}) describing the universe dominated by bulk viscous matter are obtained as
\begin{eqnarray}
3 H^2+\frac{\beta  \rho }{2 \alpha }&=&0,\label{F11}\\
2 \dot{H}+3 H^2-\frac{\beta  \Bar{p}}{2 \alpha }&=&0. \label{F22}
\end{eqnarray}

Since our focus is on late-time acceleration, we prioritize the domination of non-relativistic matter in the universe, i.e., $\gamma=1$. From Friedmann equation (\ref{F22}) and Eq. (\ref{pressure}), we derive
\begin{equation}
\Dot{H}+\frac{3}{2}H^2+\frac{3\beta\zeta}{4\alpha}H=0.
\label{H1}
\end{equation}

By substituting $\frac{1}{H} \frac{d}{dt} = \frac{d}{d \ln(a)} $ into Eq. \eqref{H1}, we derive a first-order differential equation for the Hubble parameter, expressed as
\begin{equation}
\frac{dH}{d \ln(a)}+\frac{3}{2}H+\frac{3\beta\zeta}{4\alpha}=0.
\label{H2}    
\end{equation}

Integrating Eq. (\ref{H2}), we obtain the expression for the Hubble parameter in terms of redshift as
\begin{equation}
\label{Hz}
H(z)=H_0 (1+z)^{\frac{3}{2}}+\frac{\beta \zeta}{2\alpha}\left[(1+z)^{\frac{3}{2}}-1 \right],
\end{equation}
where $H_0$ denotes the present value of the Hubble parameter. Specifically, for $\alpha=-1$, $\beta=2$, and $\zeta=0$, the solution simplifies to $H(z) = H_0 (1 + z)^{3/2}$, representing the standard matter-dominated universe.

The deceleration parameter $q$ quantifies the rate at which the expansion of the universe is slowing down or speeding up. It is defined as $q = -1 - \frac{\dot{H}}{H^2}$. If $q > 0$, the universe is decelerating, meaning it is slowing down in its expansion. Conversely, if $q < 0$, the universe is accelerating, indicating that its expansion is speeding up over time. $q = 0$ corresponds to a universe expanding at a constant rate, neither accelerating nor decelerating. Then, using Eq. (\ref{Hz}), we obtain
\begin{equation}
q(z)=-1+\frac{3 (1+z)^{3/2} (\beta  \zeta +2 \alpha  H_0)}{2 \left[(1+z)^{3/2} (\beta  \zeta +2 \alpha  H_0)-\beta  \zeta \right]}.
\end{equation}

The effective EoS parameter $\omega_{eff}$ describes the relationship between the pressure ($\Bar{p}$) and the energy density ($\rho$) of a particular component of the universe, such as DE, dark matter (DM), or normal matter. In cosmology, $\omega$ characterizes different types of energy and matter contributing to the universe's dynamics. For non-relativistic matter like dust, $\omega= 0$. Radiation, such as photons and neutrinos, has an EoS parameter $\omega = \frac{1}{3}$. A CC or vacuum energy corresponds to $\omega = -1$, indicating a constant energy density causing accelerated expansion. DE that varies over time, unlike a CC, has an EoS parameter in the range $-1 < \omega < 0$. This range suggests a dynamic form of DE affecting the universe's expansion rate. The effective EoS parameter is defined as
\begin{equation}
\label{w}
\omega_{eff}=\frac{\Bar{p}}{\rho}=\frac{\beta  \zeta }{2 \alpha  H}.    
\end{equation}

By using Eqs. (\ref{Hz}) and (\ref{w}), one can obtain
\begin{equation}
\omega_{eff}=\frac{\beta  \zeta }{(1+z)^{3/2} (\beta  \zeta +2 \alpha  H_0)-\beta  \zeta }.    
\end{equation}

\section{MCMC analysis: Data sources and methodology} \label{sec4}

In this section, we address the observational features of our bulk viscous $f(Q,L_m)$ cosmological model. The best-fit values of the model parameters $H_0$, $\alpha$, $\beta$, and $\zeta$ are calculated using the $H(z)$ and Pantheon datasets. To constrain these parameters, we apply Bayesian statistical analysis and likelihood functions, employing the Markov Chain Monte Carlo (MCMC) method implemented in the emcee Python environment \cite{emcee}. We maximize the following likelihood function to find the best-fit values of the model parameters: $\mathcal{L}\propto \exp(-\chi^2 / 2)$, where the chi-squared ($\chi^2$) statistic represents a measure of the goodness of fit between observed data and model predictions. In addition, we employ the following priors: $H_0 : [60, 80]$, $\alpha : [-1, 0]$, $\beta : [0, 5]$, and $ \zeta : [0, 5]$. Our analysis also combines samples from the $H(z)$ and Pantheon datasets. The $\chi^2$ function is presented below for various datasets:

\subsection{$H(z)$ datasets}

The Hubble parameter $H(z)$ values are often obtained using the differential age of galaxies approach. The Hubble parameter at a given redshift $z$ can be estimated using the relation $H(z) = -\frac{1}{1+z} \frac{dz}{dt}$. The term $\frac{dz}{dt}$ can be determined from observations of massive and very slowly evolving galaxies, known as Cosmic Chronometers (CC). Here, we use 31 CC data points from multiple surveys that are in the redshift range $[0.1,2]$ \cite{Jimenez:2003iv,Simon:2004tf,Stern:2009ep,Moresco:2012jh,Zhang:2012mp,Moresco:2015cya,Moresco:2016mzx,Ratsimbazafy:2017vga}. To calculate the best-fit values of the model parameters, the $\chi^2$ function is defined as follows:
\begin{equation}
    \chi_{Hz}^2 = \Delta H^T (C_{CC}^{-1})\Delta H,
\end{equation}
where $\Delta H$ is the vector indicating the difference between the observed and assumed values of $H(z)$ for each redshift dataset, and $C_{CC}$ represents the covariance matrix components representing errors associated with the observed $H(z)$ values. 

Fig. \ref{F_Hz} compares the predictions of the $f(Q, L_m)$ gravity model with bulk viscosity (red solid line) and the $\Lambda$CDM model (black dashed line) against 31 observational $H(z)$ data points (blue squares with error bars). Both models show reasonable agreement with the data, with the $f(Q, L_m)$ gravity model closely aligning with the observations, particularly at lower redshifts ($z < 1.5$), while the $\Lambda$CDM model also performs well but slightly overestimates $H(z)$ at higher redshifts. The $f(Q, L_m)$ model appears to better capture some features of the data at lower redshifts, remaining closer to the observations overall at high redshifts within the error margins. This suggests that the $f(Q, L_m)$ gravity model with bulk viscosity offers a competitive alternative to $\Lambda$CDM for explaining the evolution of $H(z)$. By minimizing the chi-square function for the $H(z)$ datasets, we can determine the best-fit values for the model parameters. The $1-\sigma$ and $2-\sigma$ likelihood contours for the model parameters $H_0$, $\alpha$, $\beta$, and $\zeta$ using the $H(z)$ datasets are shown in Fig. \ref{F_CC}. These contours illustrate the confidence regions where the true values of the parameters are expected to lie with 68\% and 95\% probability, respectively. From the analysis, the best-fit values for the model parameters are $H_0=67.42\pm 0.66\,\text{km/s/Mpc}$, $\alpha=-0.124^{+0.062}_{-0.040}$, $\beta=2.96\pm 0.86$, and $\zeta=3.08^{+0.92}_{-1.0}$. It is noted that the bulk viscosity coefficient $\zeta$ is expressed in Pascal-seconds (Pa·s) in the SI system. In the Planck system, however, the units of $\zeta$ are expressed in cubic meters (m³).

\subsection{Pantheon datasets}

The Pantheon dataset, which includes 1048 SNe Ia data points, was recently released. The dataset is comprised of contributions from the Pan-STARRS1 Medium Deep Survey, the Sloan Digital Sky Survey (SDSS), the Supernova Legacy Survey (SNLS), several low-redshift surveys, and the Hubble Space Telescope (HST) surveys. Scolnic et al. \cite{Scolnic/2018} compiled the Pantheon dataset, which includes 1048 SNe Ia spanning a redshift range of $z \in [0.01, 2.3]$. For the Pantheon dataset, the $\chi^2$ function is defined as follows:
\begin{equation}
\chi^2_{SNe}=\sum_{i,j=1}^{1048}\Delta\mu_{i}\left(C^{-1}_{SNe}\right)_{ij}\Delta\mu_{j},
\end{equation}
where $C_{SNe}$ is the covariance matrix \cite{Scolnic/2018}, and $\Delta\mu_{i}=\mu^{th}(z_i,H_0,\alpha,\beta,\zeta)-\mu_i^{obs}$ represents the difference between the observed distance modulus from cosmic observations and the corresponding theoretical values predicted by the model. In addition, the theoretical distance modulus is expressed as $\mu^{th}(z)= 5log_{10} d_{L}(z)/Mpc+25$, where, in the case of a spatially flat universe, $d_{L}$ is the luminosity distance given by
\begin{equation}
d_L(z) = c(1+z)\int_0^z \frac{dy}{H(y)}.
\end{equation}

Here, $c$ denotes the speed of light.

In a similar manner, the $1-\sigma$ and $2-\sigma$ likelihood contours for the model parameters $H_0$, $\alpha$, $\beta$, and $\zeta$ using the Pantheon datasets are depicted in Fig. \ref{F_SN}. From the analysis, the best-fit values for the model parameters are $H_0=67.21\pm 0.77\,\text{km/s/Mpc}$, $\alpha=-0.116^{+0.063}_{-0.038}$, $\beta=2.94\pm 0.89$, and $\zeta=3.10^{+0.93}_{-1.0}$.

\subsection{$H(z)$+Pantheon datasets}

The $\chi^2$ function for the $H(z)$+Pantheon datasets is expressed as
\begin{equation}
    \chi^{2}_{total} = \chi^{2}_{Hz} + \chi^{2}_{SNe}.
\end{equation}

Moreover, Fig. \ref{F_CC+SN} shows the $1-\sigma$ and $2-\sigma$ likelihood contours for the model parameters $H_0$, $\alpha$, $\beta$, and $\zeta$ using the combined $H(z)+$Pantheon datasets. The analysis yields the best-fit values for the model parameters: $H_0=67.22\pm 0.55\,\text{km/s/Mpc}$, $\alpha=-0.122^{+0.065}_{-0.040}$, $\beta=2.95^{+0.96}_{-0.86}$, and $\zeta=3.15^{+0.90}_{-1.1}$. These results from all datasets indicate that the value of $H_0$ is consistent with recent measurements from various cosmological observations, suggesting a moderate expansion rate \cite{Planck/2014,Planck/2020}.

Next, we will discuss the behavior of cosmological parameters that describe the evolutionary phases of the universe, based on the extracted values of the model parameters from various observational datasets, such as the $H(z)$, Pantheon, and $H(z)$+Pantheon datasets. 

\begin{widetext}

\begin{figure}[H]
\centering
\includegraphics[width=18cm,height=6cm]{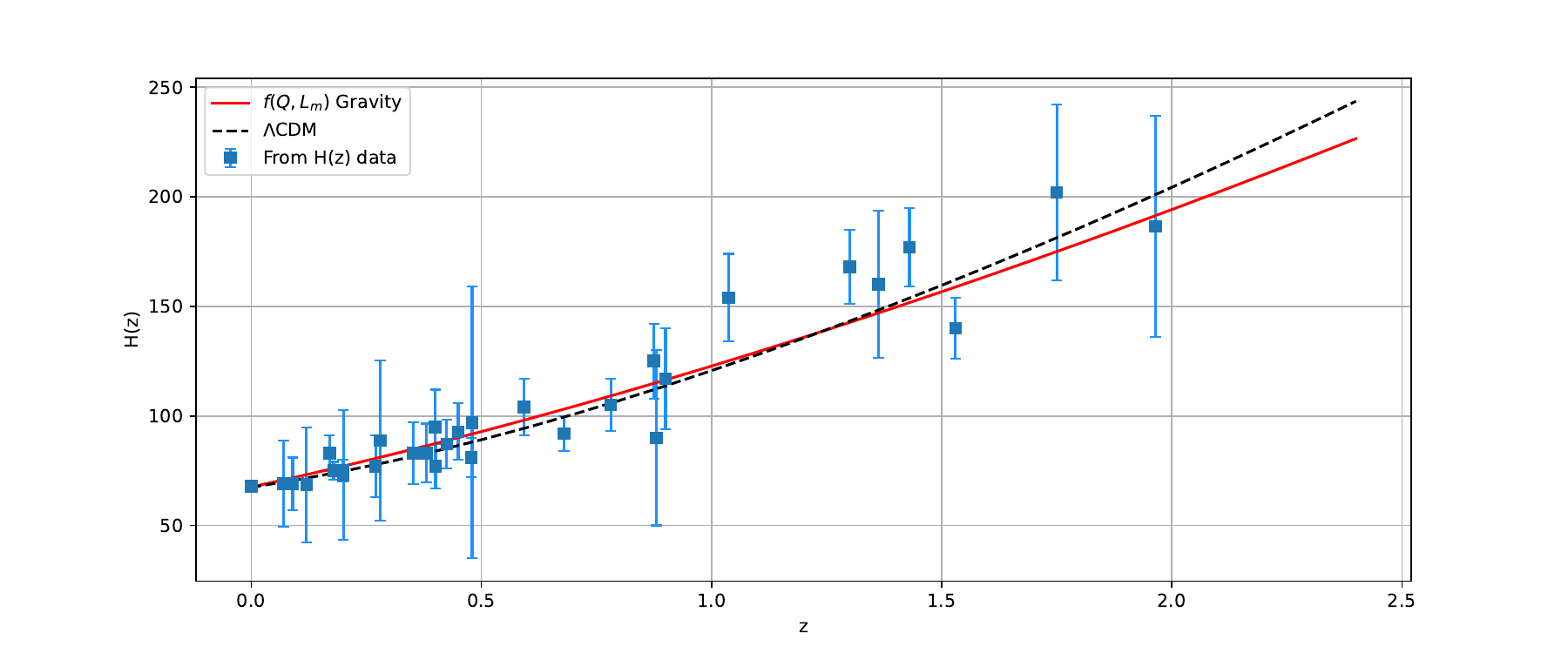}
\caption{A comparison of the predictions from the $f(Q, L_m)$ gravity model with bulk viscosity and the $\Lambda$CDM model, using observational data of $H(z)$ derived from 31 data points.}
\label{F_Hz}
\end{figure}

\begin{figure}[H]
\centering
\includegraphics[scale=0.7]{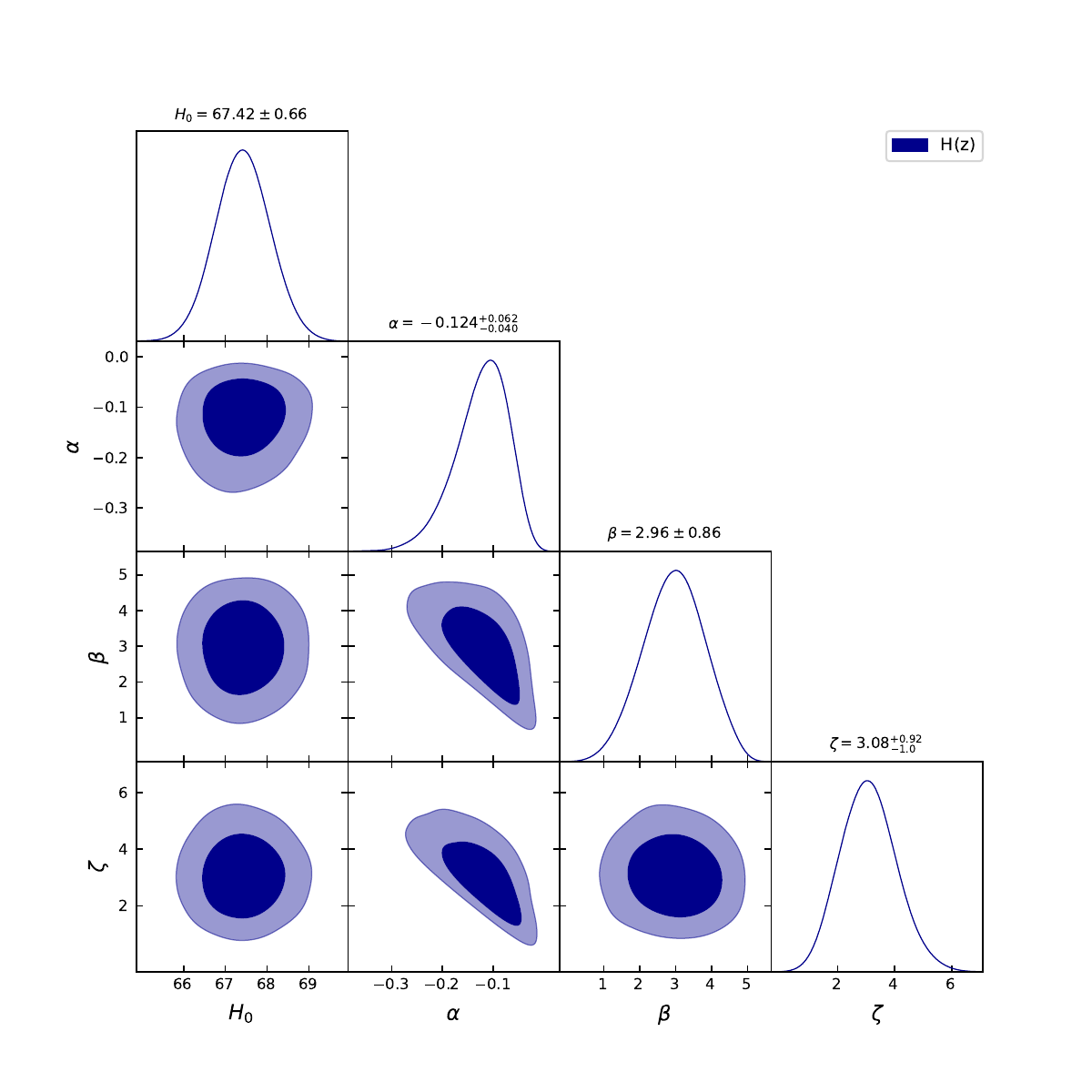}
\caption{The likelihood contours of $1-\sigma$ and $2-\sigma$ for the model parameters $H_0$, $\alpha$, $\beta$, and $\zeta$, obtained from the $H(z)$ datasets.}
\label{F_CC}
\end{figure}

\begin{figure}[H]
\centering
\includegraphics[scale=0.7]{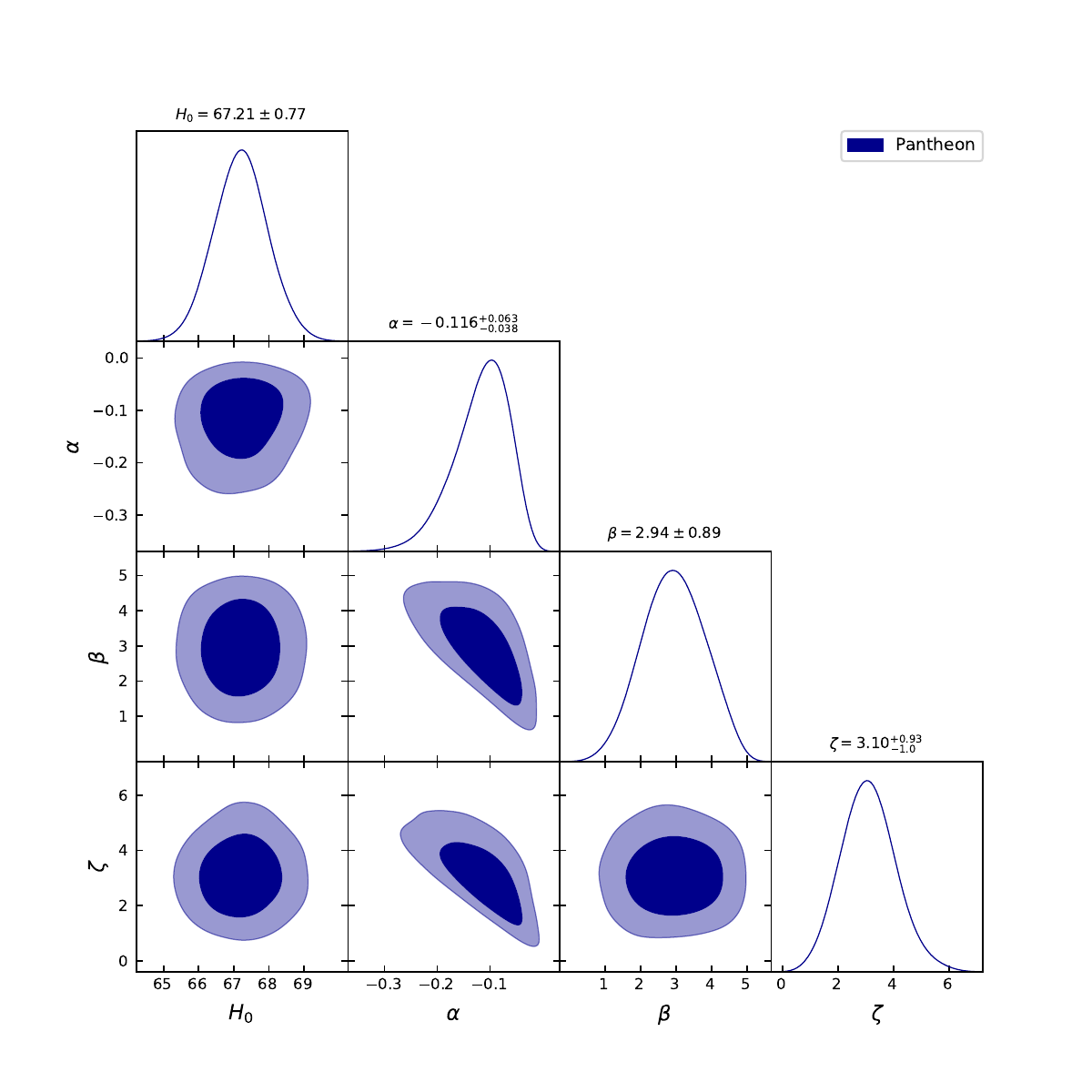}
\caption{The likelihood contours of $1-\sigma$ and $2-\sigma$ for the model parameters $H_0$, $\alpha$, $\beta$, and $\zeta$, obtained from the Pantheon datasets.}
\label{F_SN}
\end{figure}

\begin{figure}[H]
\centering
\includegraphics[scale=0.7]{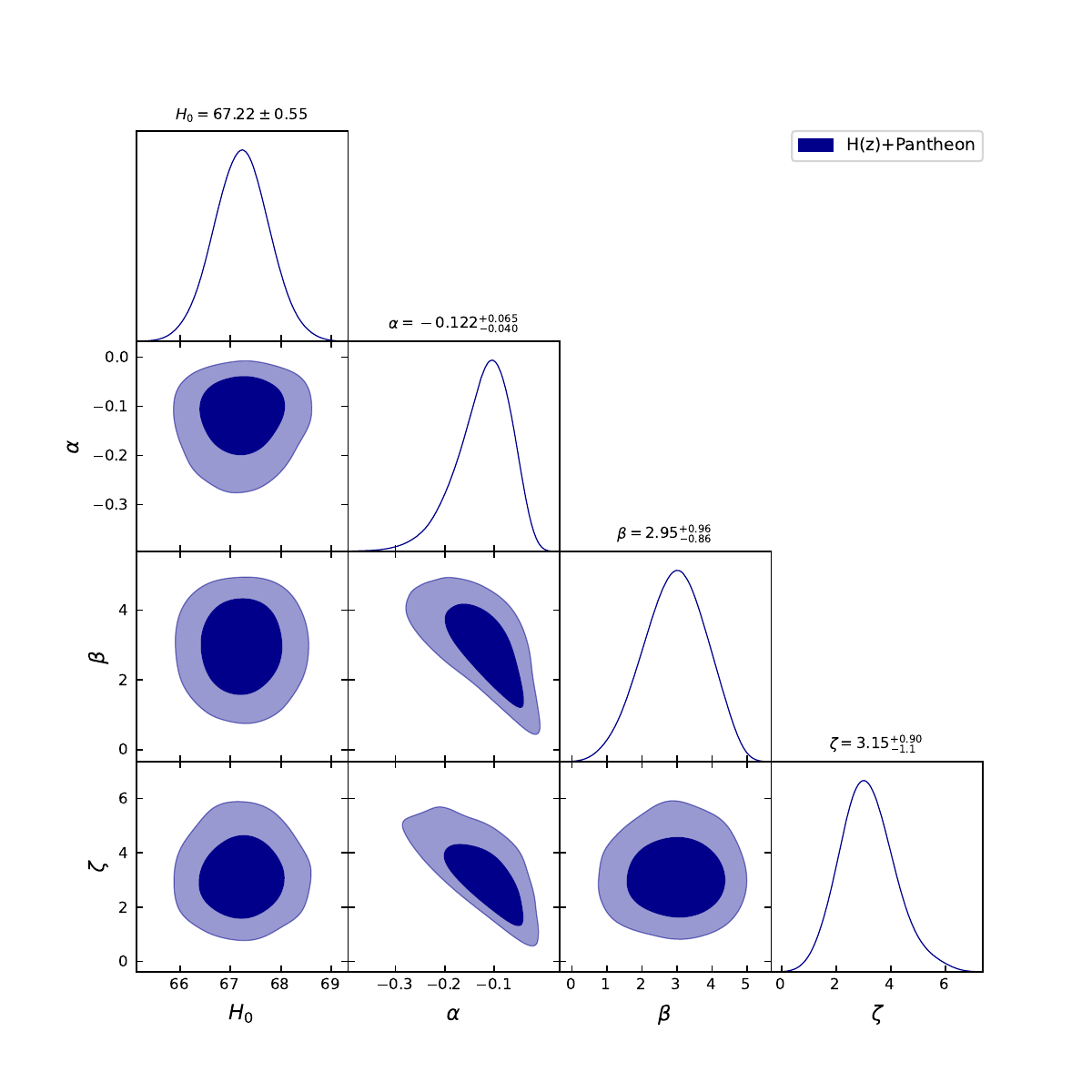}
\caption{The likelihood contours of $1-\sigma$ and $2-\sigma$ for the model parameters $H_0$, $\alpha$, $\beta$, and $\zeta$, obtained from the $H(z)$+Pantheon datasets.}
\label{F_CC+SN}
\end{figure}

\end{widetext}

\begin{figure}[h]
\centering
\includegraphics[scale=0.7]{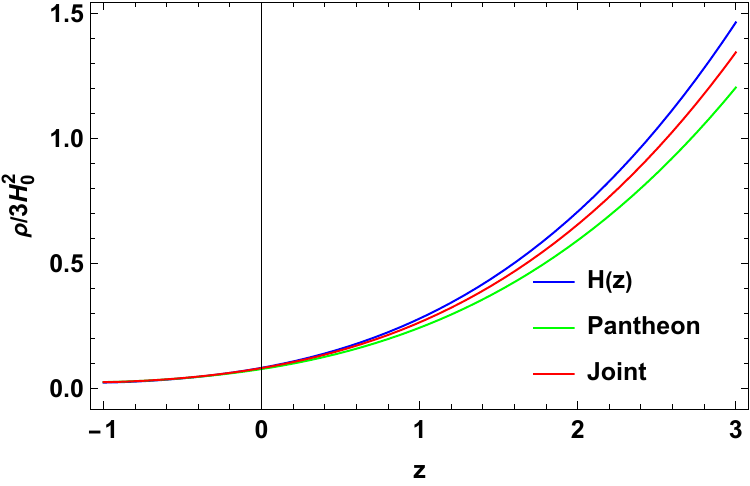}
\caption{Evolution of energy density versus redshift.}
\label{F_rho}
\end{figure}

\begin{figure}[h]
\centering
\includegraphics[scale=0.7]{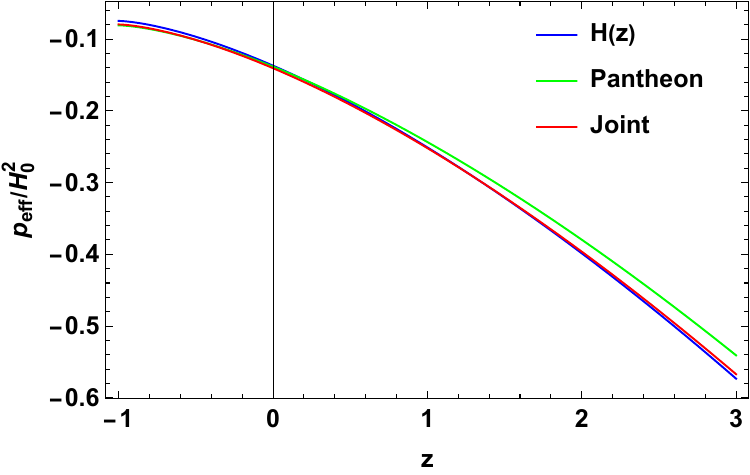}
\caption{Evolution of effective pressure versus redshift.}
\label{F_p}
\end{figure}

Figs. \ref{F_rho} and \ref{F_p} present the evolution of the energy density and effective pressure, respectively. The energy density demonstrates the expected positive behavior, aligning with theoretical predictions for a universe dominated by DM and DE. Meanwhile, the pressure, influenced by the viscosity coefficient, exhibits negative behavior across the entire domain. This negative pressure, often associated with DE, suggests a repulsive force that drives the accelerated expansion of the universe. The inclusion of the viscosity coefficient appears to enhance this effect, indicating that viscous effects in the cosmic fluid play a significant role in the dynamics of the universe's expansion. These findings support the hypothesis that viscosity in the cosmic content contributes to the overall accelerated expansion, providing a potential explanation for the observed accelerated growth of the universe.

\begin{figure}[h]
\centering
\includegraphics[scale=0.7]{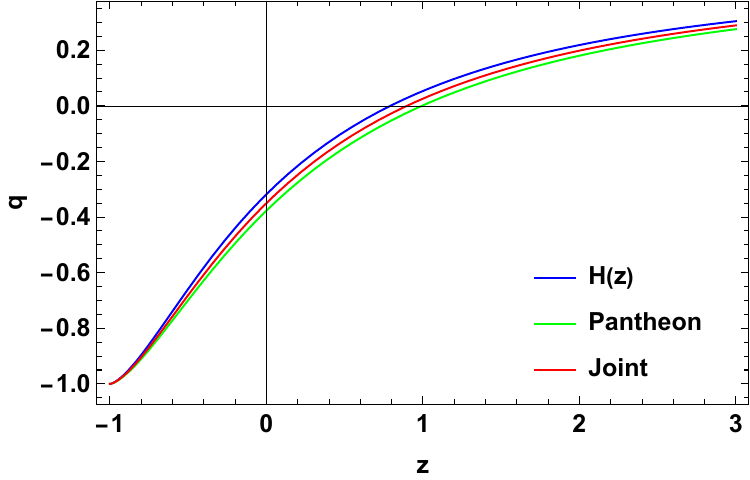}
\caption{Evolution of deceleration parameter versus redshift.}
\label{F_q}
\end{figure}

\begin{figure}[h]
\centering
\includegraphics[scale=0.7]{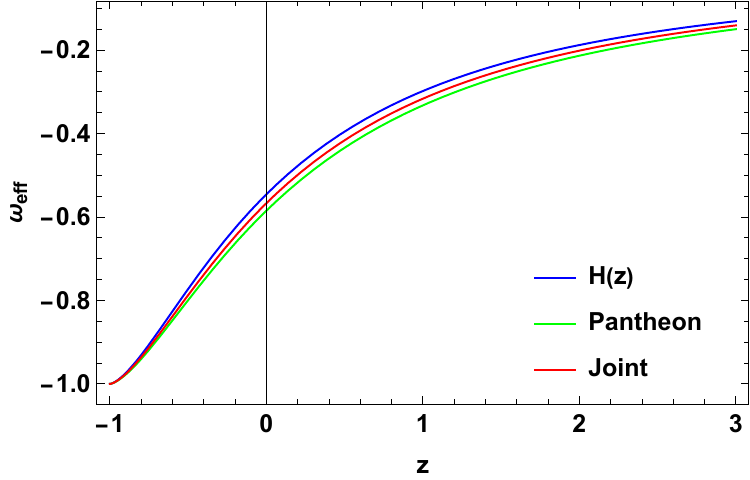}
\caption{Evolution of effective EoS parameter versus redshift.}
\label{F_w}
\end{figure}

From Fig. \ref{F_q}, it is evident that the deceleration parameter indicates a transition from a decelerated phase ($q > 0$) to an accelerated phase ($q < 0$) of the universe's expansion for the constrained values of the model parameters. The transition redshift is $z_t \approx 0.80$, $z_t \approx 0.99$, and $z_t \approx 0.90$ according to the $H(z)$, Pantheon, and $H(z)$+Pantheon datasets, respectively. The current value of the deceleration parameter is $q_0 = -0.31$, $q_0 = -0.38$, and $q_0 = -0.35$ for the respective datasets \cite{Hernandez,Basilakos,Roman,Jesus,Cunha}. Moreover, the effective EoS parameter depicted in Fig. \ref{F_w} indicates that the cosmic viscous fluid exhibits characteristics similar to quintessence DE ($-1<\omega_0<-\frac{1}{3}$). The current values of the EoS parameter according to the $H(z)$, Pantheon, and $H(z)$+Pantheon datasets are $\omega_0 = -0.54$, $\omega_0 = -0.59$, and $\omega_0 = -0.56$, respectively \cite{Gruber,Y_Myrzakulov}.

\section{$Om(z)$ diagnostics} \label{sec5}

In the literature, statefinder parameters and $Om(z)$ diagnostic analysis are commonly employed to differentiate between DE models \cite{Sahni/2003,Sahni/2008}. To understand cosmological models, the Hubble, deceleration, and EoS parameters are crucial. DE models are known to produce a positive Hubble parameter and a negative deceleration parameter. Therefore, $H$ and $q$ are not sufficient to effectively differentiate between different DE models. Hence, the $Om(z)$ diagnostic analysis plays a crucial role in this type of analysis and helps identify deviations from $\Lambda$CDM by examining the evolution of the Hubble parameter with redshift. For a spatially flat universe, it is defined as
\begin{equation}
Om\left( z\right) =\frac{\left( \frac{H\left( z\right) }{H_{0}}\right) ^{2}-1%
}{\left( 1+z\right) ^{3}-1}.    
\end{equation}

\begin{figure}[h]
\centering
\includegraphics[scale=0.7]{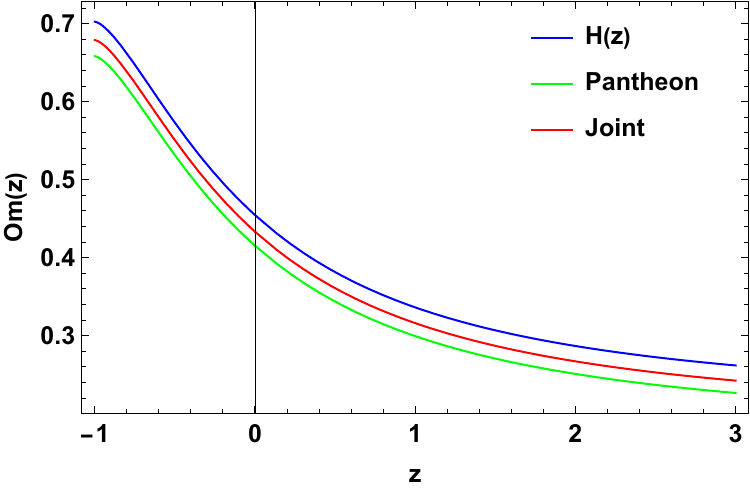}
\caption{Evolution of $Om(z)$ diagnostic versus redshift.}
\label{F_Om}
\end{figure}

Understanding these diagnostics not only aids in characterizing DE but also in predicting the future behavior of the universe's expansion, making them indispensable tools in modern cosmology. A positive, negative, or constant slope of $Om(z)$ indicates different DE models: phantom ($\omega < -1$), quintessence ($\omega > -1$), and $\Lambda$CDM ($\omega = -1$), respectively. Based on Fig. \ref{F_Om},  we can see that the $Om(z)$ diagnostic, for the constrained values of the model parameters, has a negative slope across the entire domain. Thus, we can deduce that the $f(Q, L_m)$ model dominated by bulk viscous matter behaves like a quintessence, and this is supported by the behavior of the EoS parameter.

\section{Conclusions}\label{sec6}

From a hydrodynamic point of view, it makes sense to include viscosity coefficients in cosmic matter content, since the perfect fluid assumption is just an abstraction. In this study, we investigated the role of bulk viscosity in driving late-time cosmic acceleration within an extended class of theories based on $f(Q, L_m)$ gravity \cite{Hazarika/2024}. This theory expands upon $f(Q)$ gravity by non-minimally coupling the non-metricity $Q$ with the matter Lagrangian $L_{m}$, analogous to the $f(Q,T)$ theory where $T$ is replaced by $L_{m}$. Similar to the trace-curvature couplings in energy-momentum tensor theories, the $f(Q, L_{m})$ theory's coupling between $Q$ and $L_{m}$ results in the non-conservation of the matter energy-momentum tensor. Here, We examined the function $f(Q, L_m) = \alpha Q + \beta L_m$, where $\alpha$ and $\beta$ are arbitrary parameters \cite{Xu/2019,Xu/2020}. The effective equation of state, $\bar{p} = p - 3\zeta H $, corresponds to the Einstein case value, incorporating the proportionality constant $\zeta$ used in Einstein's theory \cite{Brevik/2012,Brevik/2013,Brevik/2005}. We derived the exact solution for our bulk viscous matter-dominated $f(Q, L_m)$ model, focusing on the domination of non-relativistic matter in the universe. Subsequently, we employed various observational datasets, including $H(z)$, Pantheon, and their combination, to constrain the model parameters $H_0$, $ \alpha$, $\beta$, and $\zeta$. The obtained best-fit values are shown in Figs. \ref{F_CC}, \ref{F_SN}, and \ref{F_CC+SN}, respectively.

Our analysis has yielded several significant findings regarding the evolution and characteristics of the cosmic fluid under the influence of bulk viscosity:
\begin{itemize}
    \item \textbf{Energy dynamics}: Energy density aligns with expectations for DM and DE dominance, while viscosity-induced negative pressure suggests accelerated expansion driven by a repulsive force.
    \item \textbf{Deceleration to acceleration transition}: The deceleration parameter $q$ transitions from positive to negative values around redshifts $z_t \approx 0.80$, $z_t \approx 0.99$, and $z_t \approx 0.90$ for different datasets, indicating current accelerated expansion ($q_0 = -0.31, -0.38, -0.35$).
    \item \textbf{EoS parameter}: The effective $\omega_{eff}$ parameter resembles quintessence DE ($-1 < \omega_{eff} < -\frac{1}{3}$), with current values $\omega_0 = -0.54, -0.59, -0.56$ from respective datasets.
    \item \textbf{$Om(z)$ diagnostic}: $Om(z)$ shows a consistent negative slope, confirming quintessence-like behavior of the $f(Q, L_m)$ model dominated by bulk viscous matter.
\end{itemize}

The influence of bulk viscosity, particularly in late-time cosmic acceleration, is primarily inferred through its cosmological implications, such as its effect on the Hubble parameter \cite{Moresco:2015cya,Moresco:2016mzx,Ratsimbazafy:2017vga} and other cosmological observables like SNe Ia data \cite{Scolnic/2018}, BAO \cite{Blake/2011, Percival/2010,Giostri/2012}, and CMB \cite{Planck/2020}. The bulk viscosity acts similarly to quintessence-type DE, meaning its effects can be observed indirectly by analyzing deviations from the standard $\Lambda$CDM model in late-time cosmological data. To directly verify or distinguish the effect of bulk viscosity in experiments, we suggest focusing on its signature in large-scale structure formation and galaxy clustering, as bulk viscosity influences the effective EoS and, consequently, the rate of structure formation. Observational surveys, such as the upcoming Euclid \cite{Euclid} and the Vera Rubin Observatory’s LSST \cite{vera}, could provide more detailed data on how the viscosity affects cosmic structure at various redshifts. In addition, precise measurements of the growth rate of cosmic structures and redshift space distortions might offer a way to detect the impact of viscosity in large-scale surveys. As for the possibility of DM domination at earlier times within this scenario, it is important to note that the bulk viscosity in our model could mimic the effects of DM and DE, unifying their behavior under a single framework \cite{vis_DM1,vis_DM2}. If bulk viscosity behaves similarly to a fluid with DM properties at earlier times, laboratory tests might focus on the properties of DM candidates with similar phenomenological behavior. Although bulk viscosity is generally a macroscopic effect, it could be connected to microscopic physical processes such as quantum particle creation in strong gravitational fields \cite{bulk6,bulk7}. In this regard, indirect experimental tests might involve searching for signatures of bulk viscous effects in particle interactions or within high-energy astrophysical processes that could emulate similar environments, such as in the early universe. In laboratory settings, the detection of such DM candidates is a challenging task. However, experiments like direct detection searches for DM (e.g., LUX-ZEPLIN, XENONnT) \cite{lux,XENONnT} or collider experiments that explore weakly interacting massive particles (WIMPs) \cite{WIMPs} could shed light on the nature of DM, especially if these candidates can be linked to viscous-like properties or exhibit dissipative behavior. Further theoretical work would be needed to establish a direct link between the phenomenology of bulk viscosity and specific testable properties in DM searches.

\section*{Acknowledgment}
This research was funded by the Science Committee of the Ministry of Science and Higher Education of the Republic of Kazakhstan (Grant No. AP22682760).

\section*{Data Availability Statement}
This article does not introduce any new data.

\end{document}